\documentclass[preprint,aps,showpacs,endfloats]{revtex4}
\usepackage{epsfig}

\newcommand{\Be}{\begin{equation}}
\newcommand{\Ee}{\end{equation}}
\newcommand{\Bea}{\begin{eqnarray}}
\newcommand{\Eea}{\end{eqnarray}}

\begin{document}

\begin{flushright} 
LA-UR-11-11672
\end{flushright} 
\vspace*{1.5in}

\title{Physical Degrees of Freedom for Gauge Fields \\ 
and the Issue of Spin}


\author{T. Goldman \\
Theoretical Division \\
Los Alamos National Laboratory \\
Los Alamos, NM 87545 USA}

\begin{abstract}
The conflict between the physical degrees of freedom of gauge bosons and the Lorentz group irreps 
naturally used to describe their couplings to matter fields are illustrated and discussed, and applied 
to issues of linear and angular momentum. 
\end{abstract}

\maketitle


\section{Introduction}

The measurement of the quark contribution to the spin of the proton by means of 
the axial vector current has led to concerns as the small result\cite{xpt} appears to be in 
conflict with the quark picture of nucleon structure. However, the axial current is 
not identical to the spin contribution, and the naive quark model pictures do not 
include the relativistic effects that certainly occur for the light quarks. Furthermore, 
the issue has gotten entangled with the question of gauge invariant operators for 
angular momentum components. Here, I first remind the reader of the effect in 
hydrogenic atoms where the spin "loss" can be explicitly calculated for the electron 
contribution, then show the relativistic general result that, when viewed from a 
boosted frame, a fermion with a given spin orientation at rest appears to have 
lost some of that spin. Next, I show that there exists a spin-one irrep of the Lorentz 
group that carries only the physical degrees of freedom of a vector boson, and 
must have a Clebsch-Gordan relation to the usual (1/2,1/2) irrep used for gauge 
fields, thus implying that the physical and unphysical degrees of freedom of the 
latter can be unambiguously separated. I then turn to the issue of gauge invariant 
quantities and recall an old, but ignored problem for the eigenvalues of the Pauli 
Hamiltonian relative to that of Dirac. After this, I show how the usual separation 
of gauge invariant angular momentum components for QED is inconsistent since 
the components do not obey angular momentum commutation rules, but that the 
problem is solved by using the decomposition of Ref.(\cite{us}) 
which depends on identifying and separating the physical and unphysical 
parts of the gauge field in its usual representation. The same separation applies 
to momentum operators in gauge theories, and I identify a physical momentum 
which is neither the canonical nor the mechanical momentum. This decomposition 
also solves the issue of the Pauli-Dirac Hamiltonian dichotomy. Finally, I apply this 
decomposition to the originating issue of angular momentum decomposition in 
QCD and briefly discuss the relation to other proposals, some of which are now 
very similar. 

\section{Spin, Boosts and Angular Momentum}

Consider the wave function for the ground state solution to the Dirac equation 
for an electron bound to an "infinitely" heavy nucleus of charge $Z$ (as shown 
in any good text on quantum mechanics): 
\begin{equation}
\psi \propto \left[ \begin{array}{c} 1 \\
0 \\ -\imath\frac{(1-\gamma)}{Z\alpha}{\rm cos}\theta \\
\imath\frac{(1-\gamma)}{Z\alpha}{\rm sin}\theta e^{\imath\phi} \\
\end{array}
\right]
\end{equation}
where $\gamma =  \sqrt{1-Z^2\alpha^2}$ and we ignore the radial wave function 
as irrelevant to our purpose here. The normalization of this wave function is 
determined by 
\begin{equation}
\psi^{\dagger}\psi \propto 1+ [\frac{(1-\gamma)}{Z\alpha}]^2 [({\rm cos}\theta)^2 + ({\rm sin}\theta)^2]
\end{equation}
so that whether we use $\Sigma_3 = \gamma_3 \gamma_5$ or the generator 
of rotations about the 3-axis, $\sigma_{12}$, the matrix elements after angular 
integration is 
\begin{equation}
\frac{1}{1+[\frac{(1-\gamma)}{Z\alpha}]^2} < 1
\end{equation}
The difference {\em must} be made up by the electron orbital angular momentum 
and the angular momentum of the virtual photons involved in the binding of the 
system, so that  total angular momentum is conserved. 

\subsection{Boosts}

Similarly, if we start with a spin up Dirac fermion in its rest frame, and then 
boost ourselves to a different frame, either in the spin direction or in the 
plane transverse to it, we observe components no longer corresponding 
to a spin up fermion: 
\begin{equation}
\psi =  \left[ \begin{array}{c} 1 \\
0 \\ 0 \\ 0 \\
\end{array}
\right]
\rightarrow  \sqrt{\frac{E+m}{2m}} \left[ \begin{array}{c} 1 \\
0 \\ 0 \\ -(\frac{p}{E+m}) \\
\end{array}
\right]   
{\rm \bf{or}} 
\rightarrow  \sqrt{\frac{E+m}{2m}} \left[ \begin{array}{c} 1 \\
0 \\  (\frac{p}{E+m}) \\ 0 \\
\end{array}
\right]
\end{equation}
where the first corresponds to boosting transversely and preserves 
the $\Sigma_3$, the axial current matrix element, but not the matrix 
element of $\sigma_{12}$, the generator of rotations, while the 
second case corresponds to boosting in the 3-direction, which 
preserves the matrix element of $\sigma_{12}$ but not $\Sigma_3$. 

\section{Spin-One Irreps of the Lorentz Group}

A massive spin-one particle in the $(1,0)$ irrep of the Lorentz group 
with spin up has components\cite{DVA} 
\begin{equation}
\psi \propto \frac{m}{\sqrt{2}} \left[ \begin{array}{c} 1 \\
0 \\ 0 \\ 
\end{array}
\right]
\rightarrow  \frac{m}{\sqrt{2}} \left[ \begin{array}{c} \frac{p^+}{m}  \\
1 \\  \frac{m}{p^+}  \\ 
\end{array}
\right]
\end{equation}
where the arrow shows the effect of boosting along the 3-axis.  Just 
as for Weyl spinors, a parity conserving form may be constructed by 
appending to the above spinor its Wigner conjugate in the $(0,1)$ 
irrep, produced by applying the matrix 
\begin{equation}
 \left[ \begin{array}{ccc} 0 & 0 & 1 \\
0 & -1 & 0 \\  1 & 0 &  0 \\
\end{array}
\right]
\end{equation}
so that in the infinite momentum or zero mass limit one obtains the spin-one 
analog of a massless Majorana spinor in the Wigner-Weyl representation. 
This has only two spin polarizations: 
\begin{equation}
 \left[ \begin{array}{c}  1 \\
0 \\   0 \\  0 \\   0 \\ 1 \\
\end{array}
\right] {\rm and} 
 \left[ \begin{array}{c}  0 \\
0 \\   1 \\  1 \\   0 \\ 0 \\
\end{array}
\right] ,
\end{equation}
both of which are transverse, and which are just what is needed for the physical 
degrees of freedom of a photon. The coupling to fermions must then have the 
structure 
\begin{equation}
\bar{\Psi} \; \Gamma^{\xi}\phi_{\xi}\Psi = \bar{\Psi} \, \gamma^{\mu}A_{\mu}\Psi
\end{equation}
where $\xi$ runs over 6 index values and I have written the {\bf rhs} in terms of 
the conventional photon gauge field.  There must exist a set of Clebsch-Gordan 
coefficients, $\tilde{\Gamma}^{\xi}_{\mu}$, that relate the two irreps so that the 
conventional photon field is given by 
\begin{equation}
A_{\mu}  \equiv  \tilde{\Gamma}^{\xi}_{\mu}\phi_{\xi}
\end{equation}
but now with only two independent degrees of freedom. The gauge redundancy 
is only available in the conventional photon $(\frac{1}{2},\frac{1}{2})$ irrep. 

\subsection{Aside on Gauge {\bf non}-Invariance of the Pauli Hamiltonian}

Transformation of a Hamiltonian by a time-dependent unitary operator changes 
the spectrum\cite{me}: 
\begin{eqnarray} 
\imath\frac{\partial \psi^{\prime}}{\partial t} & =  &\left( UHU^{-1} - \imath U \frac{\partial U^{-1} }{\partial t}\right)  \psi^{\prime} \nonumber \\
& \equiv & H^{\prime}  \psi^{\prime} \;\;\;  {\rm with } \\
 \psi^{\prime} = U  \psi \;\; ; \;\; U & = & exp[-\imath H f(t) ]   \;\;\;  {\rm which \; produces}  \nonumber \\
< H^{\prime} >  & = & (1 + \dot{f})\Sigma_{n} |c_{n}|^2 E_{n} \nonumber  \\
& \neq &	\Sigma_{n} |c_{n}|^2 E_{n}  = \;  <  H >
\end{eqnarray}
However, this is exactly the transformation made by Ffoldy and Wouthuysen 
to transform from the Dirac to the Pauli Hamiltonian, with 
\begin{equation}
U = exp[\beta \vec{\alpha}\cdot(\vec{p}-e\vec{A})/2m ] 
\end{equation}
The problem is that 
\begin{eqnarray} 
H_{P}  & = & \left( UH_{D}U^{-1} - \imath U \frac{\partial U^{-1} }{\partial t}\right)   \nonumber \\
 &\simeq & \beta\left[ m + \frac{(\vec{p}-e\vec{A})^2}{2m} \right] -eA_{0} -\frac{e}{2m}\beta\vec{\sigma}\cdot\vec{B} \nonumber \\
 & & -\frac{\imath e}{8 m^2} \vec{\sigma} \cdot \vec{\nabla} \times \vec{E}  -\frac{e}{4 m^2} \vec{\sigma} \cdot \vec{E} \times \vec{p} 
 -\frac{e}{2 m^2} \vec{\sigma} \cdot \vec{E} 
\end{eqnarray} 
requires the last time derivative term to produce the gauge invariant 
\begin{equation}
\vec{E}  = -\vec{\nabla} A_{0}  - \frac{\partial \vec{A} }{\partial t} ,
\end{equation}
in the last line. That is, the result is only valid in Coulomb gauge where 
this last time derivative term of the vector potential term doesn't contribute 
because 
\begin{equation}
\vec{\nabla} \cdot \vec{A} = 0 \; ; \;  \vec{\sigma} \cdot \vec{\nabla} \times \frac{\partial \vec{A} }{\partial t} = 0\; ; \; 
 \vec{\sigma} \cdot  \frac{\partial \vec{A} }{\partial t} \times \vec{p} = 0 
\end{equation}
But the result must be valid in any gauge! Furthermore, if we chose this 
gauge, it would seem that the energies would not be Lorentz covariant. 
Fortunately, Manoukian\cite{man} has shown that despite appearances, 
Coulomb gauge is actually Lorentz covariant! As for the gauge invariance, we 
will solve this problem along with the gauge invariant decomposition of orbital 
angular momentum in the next section. 

\section{Gauge Invariant Canonical Angular Momentum} 

The straightforward decomposition of angular momentum in QED is not gauge invariant: 
\begin{eqnarray} 
\vec{J}_{QED} & = & \vec{S}_{e} + \vec{L}_{e} + \vec{S}_{\gamma} + \vec{L}_{\gamma} \;\;\;\;  {\rm where}   \nonumber \\
\vec{S}_{e} & = & \int d^3 x \; \psi^{\dagger} \frac{\vec{\Sigma}}{2} \psi \;\; ; \;\;
\vec{L}_{e}  =   \int  d^3 x \; \psi^{\dagger} \vec{x} \times  \frac{1}{\imath}\vec{\nabla} \psi  \nonumber \\
\vec{S}_{\gamma} & = &  \int   d^3 x \; \vec{E} \times \vec{A}  \;\; ; \;\;
 \vec{L}_{\gamma}  =   \int  d^3 x \; \vec{x} \times E^{i} \vec{\nabla}  A^{i}
\end{eqnarray} 
but the components of the commonly described gauge-invariant form 
\begin{eqnarray} 
\vec{J}_{QED} & = & \vec{S}_{e} + \vec{L}^{\prime}_{e} + \vec{J}^{\prime}_{\gamma} \;\;\;\;  {\rm where}   \nonumber \\
\vec{S}_{e} & = & \int d^3 x \; \psi^{\dagger} \frac{\vec{\Sigma}}{2} \psi \;\; ; \;\;
 \vec{L}^{\prime}_{e}  =   \int  d^3 x \; \psi^{\dagger} \vec{x} \times  \frac{1}{\imath}\vec{D} \psi  \nonumber \\
\vec{J}^{\prime}_{\gamma} & = &  \int   d^3 x  \;\; \vec{x} \times \left( \vec{E} \times \vec{B}  \right)
\end{eqnarray} 
do not obey the canonical commutation relations for angular momentum, viz. 
\begin{equation}
\left[ (\vec{x} \times \frac{1}{\imath}\vec{\nabla})_j \; , \; (\vec{x} \times \frac{1}{\imath}\vec{\nabla})_k \right]
=  \imath \epsilon_{jkl} (\vec{x} \times \frac{1}{\imath}\vec{\nabla})_l
\end{equation}
but 
\begin{equation}
\left[ (\vec{x} \times \frac{1}{\imath}(\vec{\nabla}-\imath e \vec{A}))_j \; , \; (\vec{x} \times \frac{1}{\imath}(\vec{\nabla}-\imath e \vec{A}))_k \right]
=  \imath \epsilon_{jkl} \{(\vec{x} \times \frac{1}{\imath}(\vec{\nabla}-\imath e \vec{A}))_l + e\;x_{l}\vec{x}\cdot (\vec{\nabla} \times \vec{A})\} .
\end{equation}
The extra term can be avoided if instead of the full gauge field, we define a part, $\vec{A}_{pur}$, 
such that $\vec{\nabla}\times \vec{A}_{pur} = 0$, which removes the last term. (See, e.g., 
Ref.(\cite{SD}).) Thus, we have proposed the decomposition\cite{us}
\begin{eqnarray} 
\vec{J}_{QED} & = & \vec{S}_{e} + \vec{L}^{\prime\prime}_{e} + \vec{S}^{\prime\prime}_{\gamma} 
+ \vec{L}^{\prime\prime}_{\gamma} \;\;\;\;  {\rm where}   \nonumber \\
\vec{S}_{e} & = & \int d^3 x \; \psi^{\dagger} \frac{\vec{\Sigma}}{2} \psi \;\; ; \;\;
\vec{L}^{\prime\prime}_{e}  =   \int  d^3 x \; \psi^{\dagger} \vec{x} \times  \frac{1}{\imath}\vec{D}_{pur} \psi  \nonumber \\
\vec{S}^{\prime\prime}_{\gamma} & = &  \int   d^3 x \; \vec{E} \times \vec{A}_{fys}  \;\; ; \;\;
 \vec{L}^{\prime\prime}_{\gamma}  =   \int  d^3 x \; \vec{x} \times E^{i} \vec{\nabla}  A^{i}_{fys}
\end{eqnarray} 
which are all gauge invariant because $\vec{A}_{fys}$ is, and where 
\begin{eqnarray} 
\vec{A} \equiv \vec{A}_{fys} +  \vec{A}_{pur}  & , &  \vec{D}_{pur}  \equiv  \vec{\nabla} - \imath e  \vec{A}_{pur}  \nonumber \\
\vec{\nabla} \cdot \vec{A}_{fys} = 0  & , &  \vec{\nabla} \times \vec{A}_{pur} = 0 .  \label{fyspur}
\end{eqnarray} 
Note, this is {\bf not} the same as choosing Coulomb gauge; it separates out the physical 
(transverse) component of the photon field as we showed above must be possible since 
a pure spin-one field irrep exists in the Lorentz group. If the gauge field is presented in 
any fixed gauge, the physical part may be projected out by using the constraints in Eqs.(\ref{fyspur}) 
and the pure gauge part identified as $\vec{A}_{pur} = \vec{A} - \vec{A}_{fys}$. The explicit 
construction is given by
\begin{eqnarray} 
\vec{A}_{fys}(x)  & = &  \vec{\nabla} \times \frac{1}{4\pi} \int   d^3y \;  \frac{\vec{\nabla} \times  \vec{A}(y)}{|\vec{x}-\vec{y}|}  \nonumber \\
{A^0}_{fys} & = & \int_{-\infty}^x dx_i \; (\partial^i A^0+\partial_t A^i - \partial_t {A^i}_{fys} )  \nonumber \\ 
\phi(x) & = & - \frac{1}{4\pi} \int   d^3y \;  \frac{\vec{\nabla}\cdot  \vec{A}(y)}{|\vec{x}-\vec{y}|}  + \phi_0(x) \nonumber \\
\vec{A}_{pur} & = &  -\vec{\nabla}\phi(x) \;\; ; \;\; {A^0}_{pur} = \partial_t\phi(x) \;\; ; \;\; \nabla^2\phi_0(x) = 0
\end{eqnarray} 

\subsection{Aside on Linear Momentum}

In the same way as for angular momentum, the bypassed problem of linear momentum 
can be solved. Neither the canonical momentum, nor the mechanical momentum
\begin{equation}
\vec{p} = m\stackrel{\cdot}{\vec{r}} + q \vec{A} = \frac{1}{\imath}\vec{\nabla} \;\; ; \;\; 
\vec{p} - q \vec{A} = m\stackrel{\cdot}{\vec{r}} = \frac{1}{\imath}\vec{D}
\end{equation}
satisfies both the commutation relation for linear momentum and gauge invariance. Since 
a gauge transformation only affects the time component and longitudinal part of the vector 
potential, the separation we identified above as $pur$ and $fys$  corresponds to "parallel" 
and transverse separation, hence we recognize the {\em physical} momentum as 
\begin{equation}
\vec{D}_{pur} =\vec{p} - q \vec{A}_{\parallel} =   \frac{1}{\imath}\vec{\nabla} - q \vec{A}_{\parallel}
\end{equation}
where $\vec{A}$ has been separated into parallel and perpendicular components in the 
same way as above
\begin{equation}
\vec{\nabla} \cdot \vec{A}_{\perp} = 0  \;\; ; \;\;  \vec{\nabla} \times \vec{A}_{\parallel} = 0 
\end{equation}
This also completes the solution of the problem of the gauge invariant Hamiltonian for the 
hydrogen atom, where the physical Hamiltonian is given by 
\begin{equation}
H_{fys} = H + q\partial_t \omega(x) = \frac{(\vec{p} - q\vec{\nabla}\omega - \vec{A}^c_{\perp})^2}{2m} + q\phi^c
\end{equation}
and $\omega(x)$ is the phase in the gauge change from the Coulomb values (labelled $^c$) 
for the 4-vector potential and $\vec{p} - q\vec{\nabla}\omega$ is identified as the physical 
momentum as above. 

\section{Application to QCD}

Again, the natural decomposition of spin and orbital angular momentum is not gauge invariant\cite{Ji}, 
but the gauge invariant forms do not obey the canonical commutation relations for angular 
momenta. We have proposed 
\begin{eqnarray} 
\vec{J}_{QCD} & = & \vec{S}_{q} + \vec{L}^{\prime\prime}_{q} + \vec{S}^{\prime\prime}_{g} 
+ \vec{L}^{\prime\prime}_{g} \;\;\;\;  {\rm where}   \nonumber \\
\vec{S}_{q} & = & \int d^3 x \; \psi^{\dagger} \frac{\vec{\Sigma}}{2} \psi \;\; ; \;\;
\vec{L}^{\prime\prime}_{q}  =   \int  d^3 x \; \psi^{\dagger} \vec{x} \times  \frac{1}{\imath}\vec{D}_{pur} \psi  \nonumber \\
\vec{S}^{\prime\prime}_{g} & = &  \int   d^3 x \; \vec{E} \times \vec{A}_{fys}  \;\; ; \;\;
 \vec{L}^{\prime\prime}_{g}  =   \int  d^3 x \; \vec{x} \times E^{i} \vec{\mathcal{D}}_{pur}  A^{i}_{fys}
\end{eqnarray} 
where $\vec{A}$ is now a matrix quantity; the gauge covariant derivatives
\begin{equation}
\vec{D}_{pur} =  \vec{\nabla}  - \imath g \vec{A}_{pur}   \;\; ; \;\;
\vec{\mathcal{D}}_{pur}  = \vec{\nabla} - \imath g [\vec{A}_{pur}, \;\;\;\; ]
\end{equation} 
and the defining constraints are
\begin{eqnarray} 
\vec{D}_{pur} \times \vec{A}_{pur} =  \vec{\nabla} \times \vec{A}_{pur} - \imath g \vec{A}_{pur} \times \vec{A}_{pur}  = 0  \\
\vec{\mathcal{D}}_{pur} \cdot \vec{A}_{fys} = \vec{\nabla} \cdot \vec{A}_{fys} - \imath g [{A}^{i}_{pur}, A^{i}_{fys}]  =   0
\end{eqnarray} 
Recall here that the color cross-product is not zero due to the matrix nature of $\vec{A}$. 
These constraints must now be solved perturbatively, but the effect is as one expects 
for a gauge transformation: 
\begin{equation}
\vec{ A}^{\prime}_{fys} = U  \vec{A}_{fys} U^{\dagger} \;\; ; \;\; \vec{ A}^{\prime}_{pur} = 
U  \vec{A}_{pur} U^{\dagger} - \frac{\imath}{g} U \vec{\nabla} U^{\dagger}
\end{equation}
For completeness, I also list the explicit time derivative parts of the equations:  
\begin{eqnarray} 
\partial_t { A}^{0}_{fys} & = & \partial_i A^0 +  \partial_t  ({A}^i - {A}^i_{fys})  -\imath g  [ ({A}^i - {A}^i_{fys}),  ({A}^0 - {A}^0_{fys})] \\
\partial_i { A}^{0}_{pur} & = & -  \partial_t  {A}^i_{pur}  + \imath g  [{A}^{i}_{pur}, A^{0}_{pur}] \
\end{eqnarray}

\subsection{Alternative Proposals}

Jaffe and Manohar\cite{JM} solve the problem on the light-cone, where there is only 
helicity to deal with. As cited above, Ji\cite{Ji} ignores the conflict with commutation 
relations in favor of classic gauge invariant operators, and Leader\cite{EL} takes the 
same viewpoint.  Wakamatsu\cite{Wak} proposes a different apportionment of our 
decomposition, which has two problems: It again violates the angular momentum 
commutation algebra and its frame independence actually conflicts with Lorentz 
invariance, as it is only $J$, the total angular momentum, that is covariant -- as we 
showed above, boosts reduce rest frame spin, shifting it to orbital angular momentum. 
Finally, Cho {\it et al.}\cite{CGPZ} propose a decomposition between gluons that they 
identify as "valence" or as "binding" gluons. This also appears to violate the angular 
momentum commutation algebra. 

\section{Conclusion}

The physical component of a vector gauge field can be identified in a gauge covariant fashion. 
The gauge covariant derivatives needed to extract orbital angular momentum (and mechanical 
momentum) of fermions coupled to the gauge field must include only the non-physical, pure 
gauge part of the vector gauge field so that:  Both gauge invariance and canonical commutation 
relations are satisfied in order to allow physical interpretation of the matrix elements of these 
operators.

This work was carried out in part under the auspices of the National Nuclear Security
Administration of the U.S. Department of Energy at Los Alamos National Laboratory
under Contract No. DE-AC52-06NA25396. I thank the organizers of the Cairns Pacific 
Spin 2011 meeting for arranging it in this lovely location and for inviting me to speak.

\end{document}